\documentclass[letterpaper,aps,pra,twocolumn,english,showpacs,superscriptaddress]{revtex4-1}

\usepackage{tikz}
\usepackage{pslatex}
\usepackage[T1]{fontenc}
\usepackage[latin1]{inputenc}
\usepackage{amssymb, amsmath, amsfonts, bm}
\usetikzlibrary{decorations.pathmorphing}
\usetikzlibrary{decorations.pathreplacing}
\usepackage{natbib}
\providecommand{\LyX}{L\kern-.1667em\lower.25em\hbox{Y}\kern-.125emX\@}

\definecolor{dgreen}{RGB}{50,128,128}
\usepackage{multirow}
\usepackage{graphicx}
\input{epsf}
\usepackage{babel}
\newcommand{\ket}[1]{|#1\rangle}
\newcommand{\bra}[1]{\langle #1|}
\newcommand{\mbf}[1]{\mathbf}

\begin{document}
\title{Adiabatic quantum computation with Rydberg-dressed atoms}

\author{Tyler Keating}
\affiliation{Center for Quantum Information and Control (CQuIC), University of New Mexico, Albuquerque NM 87131}
\affiliation{Department of Physics and Astronomy, University of New Mexico,
Albuquerque NM 87131}

\author{Krittika Goyal}
\altaffiliation{Krittika Goyal previously published under the name Krittika Kanjilal}
\affiliation{Center for Quantum Information and Control (CQuIC), University of New Mexico, Albuquerque NM 87131}

\affiliation{Department of Physics and Astronomy, University of New Mexico, Albuquerque NM 87131}

\author{Yuan-Yu Jau}
\affiliation{Center for Quantum Information and Control (CQuIC), University of New Mexico, Albuquerque NM 87131}
\affiliation{Sandia National Laboratories, Albuquerque, NM 87185}

\author{Grant W. Biedermann}
\affiliation{Center for Quantum Information and Control (CQuIC), University of New Mexico, Albuquerque NM 87131}
\affiliation{Department of Physics and Astronomy, University of New Mexico, Albuquerque NM 87131}
\affiliation{Sandia National Laboratories, Albuquerque, NM 87185}

\author{Andrew J. Landahl}
\affiliation{Center for Quantum Information and Control (CQuIC), University of New Mexico, Albuquerque NM 87131}
\affiliation{Department of Physics and Astronomy, University of New Mexico, Albuquerque NM 87131}
\affiliation{Sandia National Laboratories, Albuquerque, NM 87185}

\author{Ivan H. Deutsch}
\affiliation{Center for Quantum Information and Control (CQuIC), University of New Mexico, Albuquerque NM 87131}
\affiliation{Department of Physics and Astronomy, University of New Mexico, Albuquerque NM 87131}

\begin{abstract}
We study an architecture for implementing adiabatic quantum computation with trapped neutral atoms.  Ground-state atoms are dressed by laser fields in a manner conditional on the Rydberg blockade mechanism, thereby providing the requisite entangling interactions.  As a benchmark we study the performance of quantum annealing to the ground state of an Ising spin-lattice.   We model a proof-of-principle experiment in a realistic architecture, including details of the atomic implementation, with qubits encoded into the clock states of $^{133}$Cs.  Numerical simulation yield fidelities  >0.98 for up to four qubits, and implementations of 10-20 qubits are within the range of current technology.

\end{abstract}
\pacs{34.50.-s,34.10.+x}
\maketitle

In this letter, we consider a new platform for adiabatic quantum computation (AQC)~\cite{farhi} based on trapped neutral atoms whose coupling is mediated by the dipole-dipole interactions of Rydberg states.  An algorithm is implemented by continuous transformation of the Hamiltonian from an initial form whose ground state is easy to prepare to the final form whose ground state encodes the output of the algorithm.  If the energy gap between the ground and excited states is sufficiently large, the transition from initial to final Hamiltonian can be accomplished efficiently.  AQC is particularly attractive because the existence of an energy gap can make the system inherently robust to certain types of errors.

In contrast to quantum circuit implementations where atoms are excited to the Rydberg state with a resonant $\pi$-pulse~\cite{jaksch2000, saffmannature, grangier}, here we base our proposal on off-resonant Rydberg dressing of the atomic ground state, studied previously in the context of dipolar gases~\cite{rolston, pohl}.  This leads to an entangling mechanism that is more compatible with AQC, where interactions are always on and can be continuously changed to transfer from the initial to final Hamiltonian.  Such adiabatic evolution has been employed in recent cold atom/ion experiments to study quantum simulations of Ising models~\cite{Kim2010, spinchain, Bollinger12}.   As a specific example, we will show how our architecture can be used to implement ``quantum annealing'' (QA) in an Ising spin-lattice~\cite{QA} to solve an instance of the quadratic unconstrained binary optimization (QUBO) problem.   We will model the physics of its implementation to benchmark the performance of a proof-of-principle realization for a few qubits with nearest-neighbor interactions. 

The goal of the QUBO problem is to find the $N$-tuple of binary variables,
$\vec{x}=(x_1, x_2, \ldots, x_N)$, $x_i\in\{0,1\}$, that minimizes the function
\begin{equation}
\label{fz}
 f(\vec{x})= \sum_{i=1}^{N}{h_i x_i} + \sum_{i,j=1}^{N}{J_{ij} x_i x_j}.
\end{equation}
This is  equivalent to solving for the ground state of a generic Ising model, a problem that is generally NP-hard \cite{barahona}.  Nonetheless, specific instantiations of this problem map onto a variety of satisfiability and related algorithms which are tractable, and thus provide useful testbeds for the AQC architecture \cite{Karimi2012}.  Moreover, because the algorithm can be NP-hard, it is important to have multiple architectures (ion, superconductors, Rydberg atoms, ground-state atoms, etc.) in which to cross-verify the solution~\cite{Leibfried2010}.

To map QUBO to a QA algorithm, each binary variable is replaced by a projector acting on a qubit, $x_i \Rightarrow (\mathbb{I}+\sigma_z^{(i)})/2$, where the Pauli matrices are defined as usual on the qubit pseudospin, $\ket{0}=\ket{\uparrow}, \ket{1}=\ket{\downarrow}$. The solution to QUBO  maps onto finding the ground state of the ``problem Hamiltonian,'' $H_P$, in the Ising form
\begin{equation}
 H_P= \sum_{i=1}^{N}\tilde{h}_i \sigma_z^{(i)} + \sum_{i,j=1}^{N}\tilde{J}_{ij} \sigma_z^{(i)}\otimes\sigma_z^{(j)},
\label{HPn}
\end{equation}
where $\tilde{J}_{ij}= J_{ij}/4$ and $\tilde{h}_i = h_i/2 + \sum_j \tilde{J}_{ij}$.  Since the Hamiltonian commutes with all $\sigma_z^{(i)}$, the ground state is one of the computational basis states, which can be read out directly.

As a benchmark for performance of this architecture, we will study a class of Ising problems corresponding to a one-dimensional spin chain with symmetric interactions, $J_{{<}ij{>}}=J$, where ${<}ij{>}$ denotes nearest neighbors.  We choose the values $h_i$ to be equally spaced and less than $J$, $h_i =i\delta E$ with $N\delta E<J$ for $N$ qubits.  The solution to this problem is trivial; minimization is achieved with the state $\ket{1010\cdots10}$ for even $N$ or $\ket{0101\cdots10}$ for odd $N$, i.e., the bits alternate between $1$ and $0$ and the final bit is $0$.  Further, the gap between the ground and first excited states scales as $N^{-1}$, so the necessary evolution time to maintain adiabaticity grows linearly.  We consider this example only as a proof-of-principle of the method that can be modeled numerically for a few qubits and  address the critical issue of decoherence. In practice, we can accommodate more complex Ising problems on more general graphs, as we will detail later; in particular, a two-dimensional lattice would be a straightforward but NP-hard generalization \cite{barahona}, and would require no qualitative changes in the protocol described here.

To implement this test-bed algorithm in a neutral-atom system, we consider cesium atoms with qubits encoded in two  hyperfine magnetic sublevels in the ground-electronic state  of alkali-metal atoms, e.~g., the ``clock states'' of  $6S_{1/2}$ $^{133}$Cs: $\ket{0}=\ket{F=4,M_F=0}$,  $\ket{1}=\ket{F=3,M_F=0}$.   The atoms can be trapped in tightly focused optical tweezers with interatomic spacings on the order of 10 $\mu$m, thereby allowing individual addressing of qubits, similar to that already achieved in other neutral atom~\cite{Saffman2012} and ion trap~\cite{Kim2010b} experiments.  Arbitrary single qubit Hamiltonians of the form $H = \mathbf{B}\cdot\mathbf{\sigma}$ can be achieved with stimulated two-photon Raman transitions in the standard manner, with negligible photon scattering over the duration of the evolution for sufficient detuning and intensity of the lasers.  The last critical ingredient is the coupling matrix of pairwise interactions, $J_{ij}$.

To generate interactions between widely separated neutral atoms, we dress the ground state using the Rydberg blockade mechanism as discussed by Johnson and Rolston~\cite{rolston}.  The key idea is to induce a light shift (LS) on the atoms that depends on the dipole blockade.  Consider a simple model of two-level atoms, with ground-state $\ket{g}$ and excited Rydberg state $\ket{r}$,  and for interatomic separations such that the Rydberg blockade is perfect.  Double occupation of  $\ket{rr}$ is suppressed, and the laser couples the ground $\ket{gg}$ state to the excited ``bright state'' $\left( \ket{rg} + e^{i \Delta \phi} \ket{gr} \right)/\sqrt{2}$, an entangled state, where $\Delta \phi$ is the phase difference of the lasers at the positions of the atoms.  The residual difference between the LS of blockaded and noninteracting  atomic pairs is the desired coupling constant,
\begin{equation}
\label{jform}
J = E^{(2)}_{LS}- 2 E^{(1)}_{LS} \approx \frac{1}{2}\left(\Delta_r+\sqrt{\Delta_r^2+2\Omega_r^2}-2\sqrt{\Delta_r^2+\Omega_r^2},
\right).
\end{equation}
where  $\Delta_r = \omega_L -\omega_{gr}$ is the laser detuning and $\Omega_r$ is the Rabi frequency of the laser coupling the ground to Rydberg state.  Here $2 E^{(1)}_{LS}$ is the noninteracting component of the two-atom LS, which can be removed via single-atom addressed Raman lasers. When $\left|\Omega_r/\Delta_r \right| \ll 1$, $J\approx -\Omega_r^4/(8 \Delta_r^3)$.  Note, for a perfect blockade, the interaction strength is independent of the motional phase, $e^{i \Delta \phi}$, as long as the detuning is large compared to the Doppler width.  

For application to quantum computing, we require this interaction to be conditional on the states of the qubits. For the Ising problem, we seek to implement the pairwise coupling $\tilde{J}_{ij} \sigma_z^{(i)}\otimes \sigma_z^{(j)}$.  For two neighboring  atoms, we can achieve this when the detuning of the Rydberg laser is small compared with the ground-state hyperfine splitting (9.2 GHz for $^{133}$Cs) so that the LS is negligible for all but a given computational state of the targeted qubits, $\ket{x_1 x_2}$, with $x \in \{0,1\}$.  As above, this two-atom ground state is dressed through off-resonant coupling to the bright state $\left(\ket{r x_2}+e^{i \Delta \phi} \ket{x_1 r}\right)/\sqrt{2}$, and the effective interaction Hamiltonian is $H_{int} \approx J\ket{\tilde{x}_1 \tilde{x}_2}\bra{\tilde{x}_1 \tilde{x}_2}$, where $\ket{\tilde{x}_1 \tilde{x}_2}$ is the Rydberg-dressed ground state.  Up to single qubit terms (that can be compensated by individually addressed atomic LS), $H_{int} \Rightarrow \pm (J/4)\sigma_z \otimes \sigma_z$.  The positive/antiferromagnetic (negative/ferromagnetic) sign is achieved when $x_1=x_2$ $(x_1 \ne x_2)$.  The ability to choose the signs of the elements of $\tilde{J}_{ij}$ provides extra flexibility in this platform, even if the sign of the physical coupling is fixed in the dressing interaction. These concepts are illustrated in Fig. 1.

\begin{figure}[b]
\includegraphics[scale=0.63]{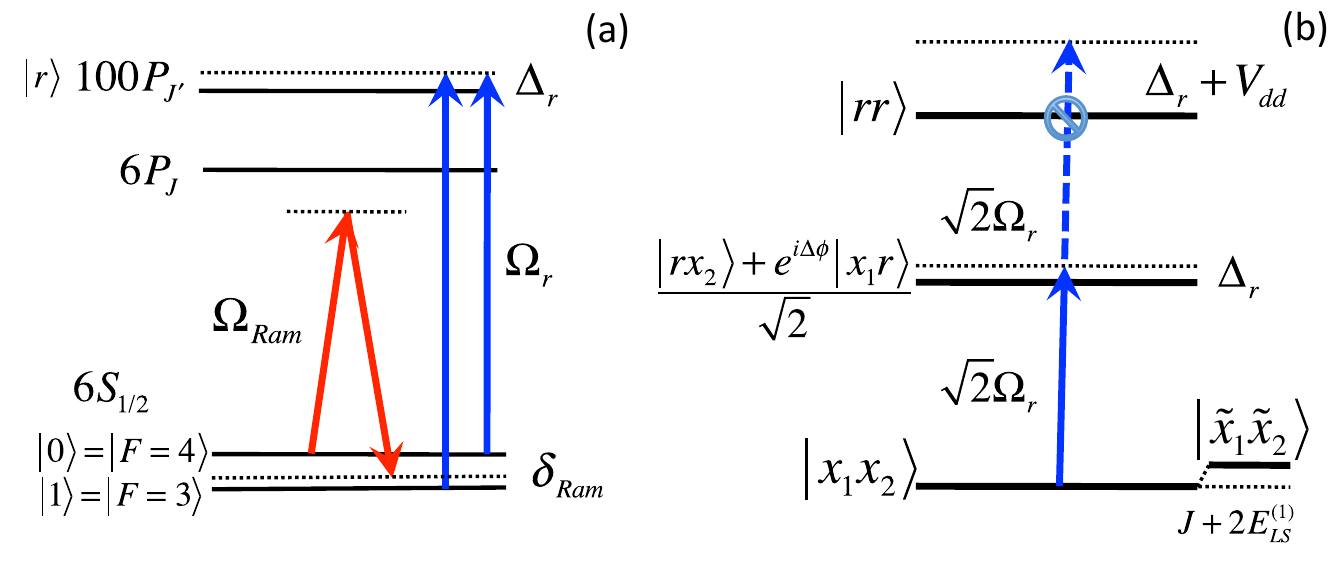}
\caption{\label{fig:cslevels}(Color online) (a) Qubits are encoded in the hyperfine clock states, controlled by stimulated Raman lasers at Rabi frequency $\Omega_{Ram}$.  Interactions between qubits are mediated by off-resonant excitation near a highly excited Rydberg state $\ket{100P_{J'}}$, with Rabi frequency $\Omega_r$, tuned to dress either $\ket{0}$ or $\ket{1}$.  (b) For two atoms,  the logical state $\ket{x_1 x_2}$, $x \in\{0,1\}$, is dressed into $\ket{\tilde{x}_1 \tilde{x}_2}$ by the bright state with one atom in the Rydberg level; the doubly excited state is blockaded by the dipole-dipole shift $V_{dd}$.  The result is a two-atom light shift with an entangling component $J$ and bare light shift $2 E^{(1)}_{LS}$.}
\end{figure}

One fundamental limitation on the fidelity of operation is the scattering of photons due to excitation of the Rydberg state at a rate $\gamma_r = N_r \Gamma_r$, where $N_r$ is the population in the Rydberg state and $\Gamma_r/2\pi$ is its linewidth.  Scattering of trap-laser photons can be made negligible with blue detuned ``bottle traps''~\cite{bottlebeam}.  While $\Gamma_r\propto n^{-3}$ points to larger principal quantum numbers, a variety of practical considerations limits the value of $n$, including the linewidth of the Rydberg excitation laser, the sensitivity of the Rydberg state to ambient fields, and the sheer size of the Rydberg atom. For example, for the 100$P$ state, the quadratic Stark shift due to the weak DC electric field is approximately $-98$ GHz$/($V$/$cm$)^2$, and the radius of the atom $r \approx 0.7$ $\mu$m.  This sets a reasonable operating point for our architecture. To illustrate our protocol, we consider here the  $100 P_{3/2}$ state, for which $\Gamma_r/2\pi = 530$ Hz.  By directly dressing the ground state with the Rydberg state  using a single optical field at $\lambda=318$ nm, we avoid the strong photon scattering that arises in the conventional two-photon excitation scheme via an intermediate excited state and reduce the total photon scattering rate by a factor of 10 or more. 

A second fundamental limitation is the accuracy with which we can implement the desired Ising Hamiltonian.  Ideally, we would like to introduce only nearest-neighbor couplings.  In practice there will be additional perturbations due to the long-range dipole-dipole interaction and the strong blockade mechanism.  For our geometry, this means that there are residual next-nearest neighbor couplings and many-body effects (e.g., \cite{threebody}) when more than two atoms are close to the blockade radius. Both interaction types will add unwanted terms to our final Hamiltonian, potentially shrinking the minimum gap or even changing the final ground state if they are too large.  However, as long as these effects can be treated as a perturbation that is sufficiently small compared to the minimum energy gap, they will not interfere with the adiabaticity of evolution, and the algorithm will still give the correct answer; this sets a minimum acceptable energy gap and, by extension, constrains the size of problem that can be solved.

We must address further details of the atomic physics in order to understand the exact nature of the coupling $J$.  The qualitative discussion above holds only for a simplified model of participating atomic levels and for a perfect dipole-blockade, i.e., when the probability of simultaneously exciting two adjacent Rydberg atoms is zero.  To obtain a more accurate description we can find the dressed-state eigenvalues by diagonalizing the two-atom system in the presence of the laser field, yielding a position dependent $J(r)$~\cite{rolston}.   Outside the so-called blockade radius, the result is  $J(r)\propto r^{-k}$, where $k=3$ for the F\"{o}rster regime or $k=6$ for the Van der Waals regime.  As we are considering direct excitation to a $p$-state, there may be concern that pairs of atoms would couple to noninteracting  ``F\"{o}rster zero states'' that evade the Rydberg blockade~\cite{saffmanreview}.  Such zeros are avoided, however, in a more complete description of the electric dipole-dipole interaction (EDDI) since mixing occurs not only between $p$- and $s$-states  but also with nearby $d$-states and higher angular momentum  orbitals.  

At small $r$ the situation becomes significantly more complex.  Our description has been based on weak EDDI mixing of nearby Rydberg levels.   Under this model, inside the blockade radius, the interaction strength $J(r)$ decreases very slowly with $r$, plateauing to the value given in Eq.~(\ref{jform}) when the blockade is considered perfect~\cite{rolston}.  For short separation distances, however, this model breaks down, as the magnitude of the EDDI  grows large relative to the bare atomic level splitting.  The result is an admixture of many more nearby Rydberg states with different $nl$  quantum numbers, leading to a splitting of  the excited states into a large ``spaghetti'' of energy levels that have more ``molecular'' than ``atomic'' character~\cite{Shaffer2006}.  

\begin{figure}[t]
\includegraphics[scale=0.355]{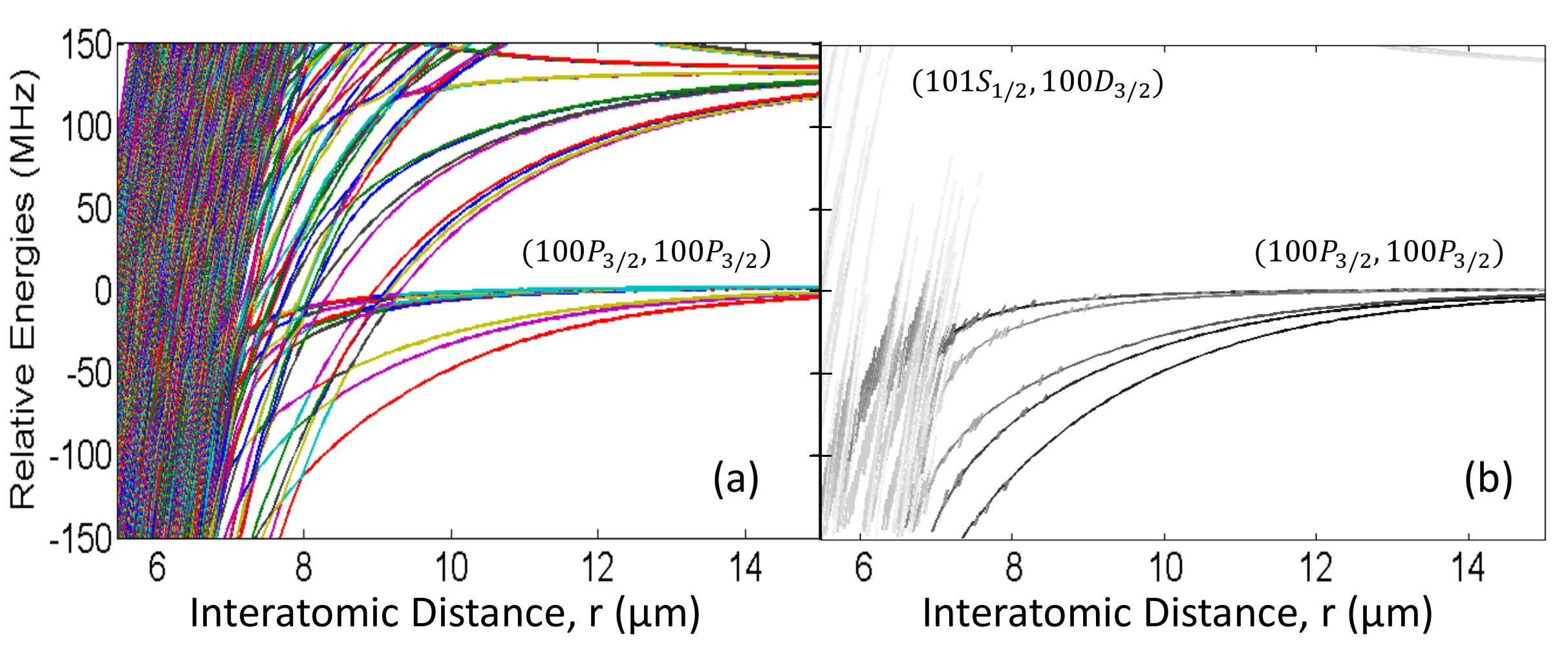}
\caption{\label{fig:spaghetti}(Color online) (a) Doubly excited levels as a function of distance between atoms that asymptote near the atomic pair $\ket{100P_{3/2}, 100P_{3/2}}$. (b) Weighting of the levels in (a) by their oscillator strengths to dress the two-atom ground states, shown in gray scale (dark = high oscillator strength).  Below $r\approx7$ $\mu$m, many levels come into resonance with $\ket{100P_{3/2}, 100P_{3/2}}$, but most have little or no oscillator strength.}
\end{figure}

Figure \ref{fig:spaghetti}a shows the portion of the doubly-excited spectrum that asymptotes at large separation near to the energy of two-atoms in the $100P_{3/2}$ state.  This spectrum is calculated using a basis of two-atom Rydberg levels $\ket{n_A l_A j_A m_{jA}}\ket{n_B l_B j_B m_{jB}}$, with $96 \le n \le 101$, $0 \le l \le 6$, and including fine structure and all magnetic sublevels, a total of $\sim$45,000 bare states.  Clearly, without proper care, this dense fan of excited levels can shift into resonance with the laser field,  potentially negating the blockade effect at the heart of our protocol.  The impact of such resonances on the dressed states will depend on the oscillator strengths for the transition that connects the two-atom ground state to these doubly-exited states via the intermediate singly-excited bright states.  Figure \ref{fig:spaghetti}b shows a plot of the spectrum, weighting each level by this oscillator strength.  We see that outside $r \approx 8$ $\mu$m there is little oscillator strength for blue detunings on the scale of interest.  Inside this radius, the situation is much less clear.  While there may be regions of small oscillator strength for shorter interatomic distances, our calculations based on the bare atomic basis fails to converge, and the results cannot be trusted.  We will restrict our attention here to $r > 8$ $\mu$m, which gives us sufficient coupling.

Given the doubly-excited spectrum, we can calculate the dressed-ground-state coupling $J(r)$.  An important consideration is the choice of detuning of the 318 nm Rydberg laser.  We seek to maximize the figure of merit $\kappa = J/\gamma_r$ so that we achieve a large gap between the ground state and excited computational states of the problem Hamiltonian, while minimizing photon scattering over the duration of the evolution.  In the simplified model that leads to Eq. (\ref{jform}), $|J|$ continues to grow as $|\Delta_r|$ decreases, while $\gamma_r$ saturates at half of the spontaneous emission rate.  This suggests that the best operating point is not one with weak dressing, but with a strong admixture of Rydberg character in the dressed-electronic-ground states that arises for small detuning.   For such strong dressing, the minimum possible detuning is set by the requirement that the gap between the dressed-ground-electronic states and the dressed-excited-Rydberg states, $\Delta E =\sqrt{2 \Omega_r^2 +\Delta_r^2}$,  is sufficient to ensure adiabatic evolution.  

\begin{figure}
\includegraphics[scale=0.7]{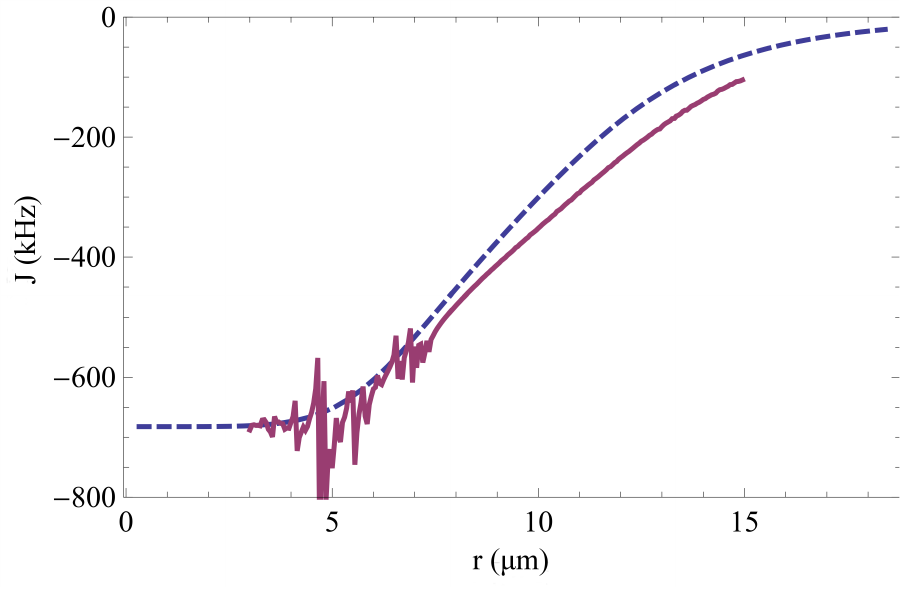}
\caption{\label{fig:Jr} (Color online) Ground-state interaction strength $J(r)$ as a function of distance between the two cesium atoms in $F=4$ clock states, for $\Omega_r=10$ MHz and $\Delta_r=8$ MHz.  Calculated using all $l\leq6$ atomic orbitals $(s,p,d,f,g,h,i$; purple solid line$)$ and more approximately using only the nearest $l\leq2$ orbitals $(s,p,d$; blue dashed line$)$.  As seen in the more exact calculation, below ~$r=8$ $\mu$m there are resonances whose exact positions cannot be predicted without taking higher-$l$ states into account.}
\end{figure}

Including the full doubly-excited spectrum shown in Fig.~\ref{fig:spaghetti}, the choice of detuning is found empirically.  For a Rydberg laser that achieves a Rabi frequency $\Omega_r = 10$ MHz, we find that a good choice of detuning is $\Delta_r = 8$ MHz.  Figure \ref{fig:Jr} shows a calculation of $J(r)$ for these parameters, and its comparison to the simplified few-level atomic model.  For  tightly trapped separated atoms,  $J(r = 8\text{ }\mu\text{m})/2\pi =-470$ kHz. At such a laser power and detuning, there is substantial dressing, with as much as $\sim$20$\%$ of Rydberg character in the dressed ground states.  The maximum photon scattering rate is $\gamma_r/2\pi \approx 100$ Hz, yielding an excellent figure of merit for AQC.  For these parameters, next-nearest-neighbor and three-body interactions for these parameters are smaller than the minimum gap for up to five atoms.  Increasing $\Delta_r$ and $r$ increases the maximum problem size that is solvable by this system at the expense of $\kappa$. This requires more runs of the experiment, but as long as the fidelity is sufficient, the probability to find the ground state can be amplified.

With these parameters we model the performance of proof-of-principle experiments to implement a simple example of QA.   The basic protocol is as follows:  We optically pump the atoms into a clock state and initialize the qubits in an eigenstate of $\sigma_x$ through the application of a  Raman-resonant $\pi/2$-pulse.  We then phase shift the Raman beam by $\pi/2$, leaving the atoms in the ground state of the beginning Hamiltonian, $H_B = -B_x \sum_{i} \sigma_x^{(i)}$, where $2 B_x=\Omega_{Ram}$ is the Raman-Rabi frequency.  This initializes the quantum register in an equal superposition of all computational basis states.  The transition from initial to final Ising Hamiltonian is achieved by decreasing  the Raman laser power while increasing the individual atoms' Raman detunings that create the local Hamiltonians $\tilde{h_i} \sigma_z^{(i)}$.  Simultaneously, we  increase the Rydberg laser power that creates the Ising Hamiltonians $J_{ij} \sigma_z^{(i)}\otimes \sigma_z^{(j)}/4$, with $J_{ij}=J$ $\forall_{i= j\pm 1}$, achieved when all atoms are arranged in an evenly spaced lattice, and negligible next-nearest interactions, as discussed above.  Note, since in our problem the coupling parameter $J_{<ij>}$ is positive while the physical $J$ is negative, we achieve the desired antiferromagnetic Ising coupling by using Rydberg laser fields that individually address the atoms, alternately dressing nearest neighbors in $\ket{0}$ and $\ket{1}$.   At the final time, the answer to the algorithm can be read out using state-dependent resonance fluorescence.  We consider here linear ramps.  More optimal time-dependent evolution can improve adiabatic following, but will depend on the specific problem.
 
We take as our parameters $B_x = J_{<ij>} = 470$ kHz, and $h_i = (i/N)118.5$ kHz for $N$ qubits, achievable with the atom-laser interactions discussed above. The ramp time is taken to be 35 $\mu$s, sufficiently long to maintain adiabatic evolution, but sufficiently short compared to the photon scattering time.  We treat spontaneous emission from the Rydberg level as effectively randomizing the magnetic spin state as the population cascades back to the electronic ground state. For practical reasons, the detection scheme does not distinguish between different magnetic sublevels in the same hyperfine subspace.  All magnetic sublevels in $F=3$ are treated as logical-1 and those in $F=4$  as logical-0.  Our simulation for two qubits, with the correct solution to ground state, $\ket{10}$, gives a fidelity of 0.997.  For larger numbers of qubits, the fidelity scales favorably.  For 3 and 4 qubits, scaling up the evolution time linearly with qubit number, we find fidelities of 0.989 and 0.990.

The performance of the neutral-atom platform for AQC depends on a combination of practical and fundamental questions.  The minimum gap between the ground state and first excited state determines the time scale for implementing the algorithm and thus the probability of spontaneous emission, the fundamental source of decoherence.  For a given problem size, the gap is constrained by $J$ arising from the Rydberg dressing, whose optimal value for a given laser power depends on the details of the atomic level structure.  We found here found that for reasonable power and detuning, we could achieve $J =470$ kHz and a fidelity of $\sim$0.99 in a proof-of-principle solution to an Ising model with $\sim$4 qubits.  Modest increases in this coupling would allow us to attain high-fidelty control with larger numbers of qubits.  However, unlike fault-tolerant universal quantum computation in the quantum circuit model, for the purpose of solving optimization problems by QA, such high fidelity is not necessary.  One requires instead the the fidelity of finding the system in the ground state be sufficiently high  that one can amplify the success probability with $k$ independent trails.  For our current parameters, this should allow us to explore the regime of 10-20 of qubits, where interesting physics beyond classical simulation is accessible.  

Finally, while this initial proof-of-principle analysis focused on nearest-neighbor Ising spin lattices, in principle this atomic architecture should allow us to explore more arbitrary connected graphs associated with a general QUBO problem.  For example, a complete bipartite graph is isomorphic to a square crosshatch of intersecting lines, where each line represents a vertex of the graph and their intersections are the edges~\cite{Karimi2012}. This could be achieved in our system by encoding logical qubits as Rydberg-coupled one dimensional spin chains~\cite{Muller2009}.  The proximity of these spin chains to one another in a designed trapping geometry would determine the edges of the graph.  Such an architecture would give substantial flexibility to explore a wide range of computationally complex Ising problems and open the door to deeper studies of QA and general AQC, as we will study in future work.

Acknowledgements: This work was supported by the Laboratory Directed Research and Development program at Sandia National Laboratories.  Sandia National Laboratories is a multi-program laboratory managed and operated by Sandia Corporation, a wholly owned subsidiary of Lockheed Martin Corporation, for the U.S. Department of Energy's  National Nuclear Security Administration under contract  DE-AC04-94AL85000.

\bibliography{RydbergAQC}

\end{document}